\documentclass[prd,nofootinbib,reprint,twocolumn,showpacs,superscriptaddress,floatfix]{revtex4-1}
\pdfoutput=1 
\usepackage{bm, color} 
\usepackage{amssymb,amsfonts,slashed,amsthm,amsmath, graphicx, soul, empheq, cancel}
\usepackage{hyperref}
\usepackage[caption=false]{subfig}

\begin{document}

\newcommand{\vev}[1]{ \left\langle {#1} \right\rangle }
\newcommand{\bra}[1]{ \langle {#1} | }
\newcommand{\ket}[1]{ | {#1} \rangle }
\newcommand{\eV}{ \ {\rm eV} }
\newcommand{\KeV}{ \ {\rm keV} }
\newcommand{\MeV}{\  {\rm MeV} }
\newcommand{\GeV}{\  {\rm GeV} }
\newcommand{\TeV}{\  {\rm TeV} }
\newcommand{\1}{\mbox{1}\hspace{-0.25em}\mbox{l}}
\newcommand{\Red}[1]{{\color{red} {#1}}}

\newcommand{\lmk}{\left(}  
\newcommand{\rmk}{\right)}
\newcommand{\lkk}{\left[}  
\newcommand{\rkk}{\right]}
\newcommand{\lhk}{\left \{ }  
\newcommand{\rhk}{\right \} }
\newcommand{\del}{\partial}  
\newcommand{\la}{\left\langle} 
\newcommand{\ra}{\right\rangle}
\newcommand{\half}{\frac{1}{2}}

\newcommand{\bea}{\begin{array}}
\newcommand{\eea}{\end{array}}
\newcommand{\beq}{\begin{eqnarray}}
\newcommand{\eeq}{\end{eqnarray}}
\newcommand{\eq}[1]{Eq.~(\ref{#1})}

\newcommand{\dd}{\mathrm{d}}
\newcommand{\Mpl}{M_{\rm Pl}}
\newcommand{\mg}{m_{3/2}}
\newcommand{\abs}[1]{\left\vert {#1} \right\vert}
\newcommand{\mphi}{m_{\phi}}
\newcommand{\Hz}{\ {\rm Hz}}
\newcommand{\for}{\quad \text{for }}
\newcommand{\Min}{\text{Min}}
\newcommand{\Max}{\text{Max}}
\newcommand{\Kahler}{K\"{a}hler }
\newcommand{\cphi}{\varphi}
\newcommand{\Tr}{\text{Tr}}
\newcommand{\diag}{{\rm diag}}
\newcommand{\rank}{{\rm rank}}

\newcommand{\SUf}{SU(3)_{\rm f}}
\newcommand{\Upq}{U(1)_{\rm PQ}}
\newcommand{\Zpq}{Z^{\rm PQ}_3}
\newcommand{\Cpq}{C_{\rm PQ}}
\newcommand{\ubar}{u^c}
\newcommand{\dbar}{d^c}
\newcommand{\ebar}{e^c}
\newcommand{\nubar}{\nu^c}
\newcommand{\Ndw}{N_{\rm DW}}
\newcommand{\Fpq}{F_{\rm PQ}}
\newcommand{\fpq}{v_{\rm PQ}}
\newcommand{\Br}{{\rm Br}}
\newcommand{\Lag}{\mathcal{L}}
\newcommand{\Lqcd}{\Lambda_{\rm QCD}}

\newcommand{\ji}{j_{\rm inf}} 
\newcommand{\jb}{j_{B-L}} 
\newcommand{\M}{M} 
\newcommand{\im}{{\rm Im} }
\newcommand{\re}{{\rm Re} }

\def\lrf#1#2{ \left(\frac{#1}{#2}\right)}
\def\lrfp#1#2#3{ \left(\frac{#1}{#2} \right)^{#3}}
\def\lrp#1#2{\left( #1 \right)^{#2}}
\def\REF#1{Ref.~\cite{#1}}
\def\SEC#1{Sec.~\ref{#1}}
\def\FIG#1{Fig.~\ref{#1}}
\def\EQ#1{Eq.~(\ref{#1})}
\def\EQS#1{Eqs.~(\ref{#1})}
\def\TEV#1{10^{#1}{\rm\,TeV}}
\def\GEV#1{10^{#1}{\rm\,GeV}}
\def\MEV#1{10^{#1}{\rm\,MeV}}
\def\KEV#1{10^{#1}{\rm\,keV}}
\def\blue#1{\textcolor{blue}{#1}}
\def\red#1{\textcolor{blue}{#1}}

\newcommand{\eff}{\Delta N_{\rm eff}}
\newcommand{\neff}{\Delta N_{\rm eff}}
\newcommand{\cc}{\Omega_\Lambda}
\newcommand{\Mpc}{\ {\rm Mpc}}
\newcommand{\Msolar}{M_\odot}

\def\sn#1{\textcolor{red}{#1}}
\def\SN#1{\textcolor{red}{[{\bf SN:} #1]}}
\def\my#1{\textcolor{blue}{#1}}
\def\MY#1{\textcolor{blue}{[{\bf MY:} #1]}}
\def\yn#1{\textcolor{magenta}{#1}}
\def\YN#1{\textcolor{magenta}{[{\bf YN:} #1]}}

\begin{flushright}
\end{flushright}

\title{
Spontaneous CP violation in Supersymmetric QCD
}

\author{Shota Nakagawa,
Yuichiro Nakai
and Yaoduo Wang
\\*[10pt]
{\it \normalsize Tsung-Dao Lee Institute, Shanghai Jiao Tong University, \\
No.~1 Lisuo Road, Pudong New Area, Shanghai, 201210, China} \\*[3pt]
{\it \normalsize School of Physics and Astronomy, Shanghai Jiao Tong University, \\
800 Dongchuan Road, Shanghai, 200240, China}\\
}

\begin{abstract}
We investigate a composite model of spontaneous CP violation based on a new supersymmetric QCD
as a solution to the strong CP problem.
The scalar components of the meson chiral superfields obtain complex vacuum expectation values to break CP symmetry spontaneously.
Then, wavefunction renormalization for the quark kinetic terms provides the Cabibbo-Kobayashi-Maskawa (CKM) phase,
while the strong CP phase $\bar{\theta}$ is protected by nonrenormalization of the superpotential and
hermiticity of the wavefunction renormalization factor.
In our model, the right-handed down-type quark multiplets are given by composite states,
enhancing their couplings to CP breaking fields, which is essential to realize the observed CKM phase.
The non-perturbative dynamics generates the scale of spontaneous CP violation hierarchically lower than the Planck scale.
We discuss potential corrections to $\bar{\theta}$ and find a viable parameter space of the model to solve the strong CP problem
without fine-tuning.
\end{abstract}

\maketitle
\flushbottom

\section{Introduction
\label{introduction}}

The Standard Model (SM) is in good agreement with the current experimental data
and provides the understanding of particles composing our visible Universe
and their interactions up to the electroweak scale. 
Nevertheless, there must exist physics beyond the SM,
one of the reasons for which is the strong CP problem that lies within the SM
(see e.g. refs.~\cite{Cheng:1987gp,Kim:2008hd,Hook:2018dlk} for reviews).
The physical strong CP phase $\bar{\theta}$ is composed of the $\theta$ parameter of
the Chern-Simons coupling of QCD and the quark mass phases.
The measurement of the neutron electric dipole moment
\cite{Abel:2020pzs}
puts a stringent upper limit on the strong CP phase, $\bar{\theta}\lesssim10^{-10}$ \cite{Pospelov:1999mv},
while the Cabibbo-Kobayashi-Maskawa (CKM) matrix contains an unsuppressed CP-violating complex phase.
The most popular approach to this strong CP problem is the Peccei-Quinn mechanism \cite{Peccei:1977hh,Peccei:1977ur}.
By introducing a global $U(1)$ symmetry which is anomalous under QCD, the strong CP phase is dynamically set to zero.
However, for the mechanism to work, the global $U(1)$ symmetry must be realized to an extraordinary high degree,
which is generally incompatible with the implication of quantum gravity \cite{Kallosh:1995hi,Banks:2010zn,Witten:2017hdv,Harlow:2018jwu}.
Another approach is spontaneous CP violation
\cite{Nelson:1983zb,Barr:1984qx,Barr:1984fh,Bento:1991ez,Barr:1993hb,Dine:1993qm,Hiller:2001qg,Hiller:2002um,Vecchi:2014hpa,Dine:2015jga,Evans:2020vil,Cherchiglia:2020kut,Cherchiglia:2021vhe,Valenti:2021rdu,Valenti:2021xjp,Fujikura:2022sot,Girmohanta:2022giy},
where CP is a symmetry of the theory so that the strong CP phase is forbidden
but the symmetry is spontaneously broken to generate the CKM phase in a way to avoid reintroducing $\bar{\theta}$.

In refs.~\cite{Hiller:2001qg,Hiller:2002um},
Hiller and Schmaltz took the advantage of beautiful features of supersymmetry (SUSY)
to deliver spontaneously broken CP into the CKM matrix without generating the strong CP phase $\bar{\theta}$.
They assumed that fields to spontaneously break CP are isolated from the quark sector
or other color charged fields at the tree-level \cite{Arkani-Hamed:1999ylh}.
In SUSY, the nonrenormalization theorem
guarantees that superpotential terms including the ${\theta}$ term do not receive radiative corrections.
On the other hand, the K$\ddot{\rm{a}}$hler potential is renormalized, and
when it is canonically normalized,
superpotential couplings are modified accordingly.
However, the hermiticity of a wave function renormalization factor protects the strong CP phase
from receiving a nonzero contribution
\cite{Ellis:1978hq,Dugan:1984qf,Holdom:1999ny}, while
the CKM phase acquires a finite contribution.

To see the Hiller-Schmaltz mechanism more explicitly, let us consider the model presented in the original papers
\cite{Hiller:2001qg, Hiller:2002um}. 
The CP-invariant superpotential is given by
\beq
W_{\rm HS}= r_{j}^{~i}D^{~p}_iF^jT_p + M_{\rm CP}T_p\bar{T}^p + \Sigma^{~j}_{i}F^i\bar{F}_j \, ,
\label{HS}
\eeq
where $D^{~p}_i$ ($i=1,2,3$ and $p=1,2,3$ denote the flavor and color indices, respectively)
is the right-handed down-type quark chiral superfield, $T_p,\bar{T}^p$ new vector-like quarks, $F^i,\bar{F}_i$
SM-singlet messengers,
and $\Sigma^{~j}_{i}$ SM-singlet chiral superfields whose complex vacuum expectation values (VEVs)
spontaneously break CP symmetry and make messenger fields massive.
The coefficients $r^{~i}_{j}$ and $M_{\rm CP}$ are real, and those of the third term are absorbed into the fields $\Sigma$.
Integrating out the heavy fields $T_p,\bar{T}^p$ and $F^i,\bar{F}_i$,
the K$\ddot{\rm{a}}$hler potential of the right-handed down-type quarks are renormalized,
and at the one-loop level, the wavefunction renormalization factor is estimated as 
\beq
\label{wavefunction}
\delta Z_{i}^{~j} \sim \frac{r_{i}^{\dagger k}\Sigma_{k}^{\dagger l}\Sigma_{l}^{~m} r^{~j}_{m}}{16\pi^2 M_{\rm CP}^2} \, .
\eeq
Canonical normalization of the right-handed down-type quarks then leads to a complex phase in the CKM matrix
without reintroducing $\bar{\theta}$.
However, Eq.~\eqref{wavefunction} indicates two problems in this mechanism. 
Firstly, the observed CKM phase requires large dimensionless couplings $|r|\sim 4\pi$
\cite{Hiller:2001qg,Hiller:2002um},
which immediately hit a Landau pole without any relief measures.
Secondly, the coincidence of scales, $|\langle\Sigma\rangle|\sim M_{\rm CP}$, is necessary.
To fully solve the strong CP problem by the Hiller-Schmaltz mechanism,
those issues need to be addressed.

From naive dimensional analysis (NDA)
\cite{Georgi:1984iz,Georgi:1986kr,Luty:1997fk,Cohen:1997rt},
large couplings $|r|\sim 4\pi$ imply that
all the fields involved in the first term of Eq.~\eqref{HS} are composite due to some new strong dynamics
and the term is given as an effective interaction after the confinement.
Then, in the present paper, 
we will consider a supersymmetric QCD to obtain the superpotential~\eqref{HS}
as its low-energy effective theory.\footnote{
For recent studies on the high-quality axion solution to the strong CP problem using supersymmetric QCD,
see refs.~\cite{Lillard:2017cwx,Lillard:2018fdt,Nakai:2021nyf,Nakagawa:2023shi}.
}
The real mass parameter of new vector-like (composite) quarks $M_{\rm CP}$ is given by
a VEV of a dynamical field, and 
meson chiral superfields $\Sigma$ obtain complex VEVs to break CP symmetry spontaneously,
so that $|\langle\Sigma\rangle|\sim M_{\rm CP}$ is naturally realized and they are 
hierarchically lower than the Planck scale due to dimensional transmutation.
We will see that the model solves the strong CP problem without fine-tuning and unnatural coincidence.
\\

The rest of the paper is organized as follows.
In \SEC{sec:model}, we introduce the model and show its low-energy description in terms of composite fields
giving the superpotential of the Hiller-Schmaltz mechanism.
A way to get the down-type quark Yukawa couplings is also explained. 
In \SEC{sec:vacuum}, the vacuum structure of the low-energy theory is analyzed,
and we will find appropriate VEVs of composite fields to generate the correct CKM phase without reintroducing
the strong CP phase.
\SEC{sec:constraint} discusses potential corrections to $\bar{\theta}$ due to SUSY breaking and
higher-dimensional operators violating CP explicitly.
\SEC{sec:discussion} is devoted to conclusions and discussions.
In particular, we will comment on the issue of domain wall production.
In Appendix \ref{app:stabilization}, we summarize details of the stabilization of vacuum moduli space,
and in Appendix \ref{app:UV}, a possible UV completion of higher-dimensional operators in the model is presented.

\section{The model
\label{sec:model}}

Let us consider a new supersymmetric $SU(5)$ gauge field theory with $N_F=6$ vector-like pairs of chiral superfields
$Q_\alpha, \bar{Q}^\alpha \, (\alpha = 1, \cdots , 6)$,
which are decomposed into (anti-)fundamental $(Q_p,\bar{Q}^p; \, p=1,2,3)$ and
singlet $(Q_i,\bar{Q}^i; \, i=4,5,6)$ representations under the ordinary color $SU(3)_c$.
The theory respects CP symmetry. 
The SM hypercharges are assigned so that some composite mesons in the low-energy effective theory
are identified as the right-handed down-type quarks, as we will see below.
To cancel the SM gauge anomaly, we introduce an $SU(5)$ singlet superfield $\Psi_i^{~p}$,
which has the same SM charges as the right-handed down-type quark
and becomes heavy after the confinement of the $SU(5)$ gauge theory.
Table \ref{tab:charge} summarizes the charge assignment of fields in the model. 
We consider the following superpotential:
\beq
W &\supset& \frac{\widetilde{m}}{3} \bar{Q}^{pn}Q_{pn}
+ \widetilde{M}_{j}^{~i} \bar{Q}^{jn}Q_{in}
+\widetilde{\iota}_j^{~i}\Psi_i^{~p}\bar{Q}^{jn}Q_{pn}\nonumber\\
&+& \frac{\widetilde{\rho}}{3}\frac{(\bar{Q}^{pn}Q_{qn})(\bar{Q}^{qn}Q_{pn})}{\mu}
+\widetilde{\lambda}_{jl}^{~ik}\frac{(\bar{Q}^{jn}Q_{in})(\bar{Q}^{ln}Q_{kn})}{\mu},\nonumber\\
\label{UV}
\eeq
where $n \, (=1,2,\cdots, 5)$ denotes the $SU(5)$ index,
$\widetilde{\iota}, \widetilde{\rho}, \widetilde{\lambda}$ are dimensionless coupling constants
and $\widetilde{m}, \widetilde{M}, \mu$ are mass parameters.
They are all real due to CP symmetry.
We assume that $\mu \lesssim M_{\rm Pl}$ and $\widetilde{m},\widetilde{M} \lesssim \Lambda \lesssim \mu$ with the dynamical scale of the theory $\Lambda$.
A possible UV completion of the higher-dimensional terms in the second line is discussed in Appendix~\ref{app:UV}.

\begin{table}[tp]
\vspace{0mm}
\centering
\begin{tabular}{c|c|c|c|c|c}
& $Q_p$ & $\bar{Q}^p$ & $Q_i$ & $\bar{Q}^i$ & $\Psi_i^{~q}$ \\
\hline
$SU(5)$ & $\square$ & $\overline{\square}$ & $\square$ & $\overline{\square}$ & $\bm 1$ \\
$SU(3)_c$ & $\square$ & $\overline{\square}$ & $\bm 1$ & $\bm 1$ & $\overline{\square}$\\
$U(1)_Y$ & $-2/15$ & $2/15$ & $1/5$ & $-1/5$ & $1/3$ \\
\end{tabular}
\vspace{1mm}
\caption{The charge assignment of fields in the model.}
\label{tab:charge}
\end{table}

\begin{table*}[tp]
\vspace{0mm}
\centering
\begin{tabular}{c|c|c|c|c|c|c|c|c|c|c}
& $\Phi$ & $\tilde{\Phi}$ & $\Sigma$ & $\Xi$ & $D$ & $\bar{T}$ & $T$ & $F$ & $\bar{F}$ & $\Psi_i^{~q}$ \\
\hline
$SU(3)_c$  & ${\bm 1}$ &  ${\rm {\bf Ad}}$ & $\bm 1$ & $\square$ & $\overline{\square}$ & $\overline{\square}$ & $\square$  & $\bm 1$ & $\bm 1$ & $\overline{\square}$ \\
$U(1)_Y$ & $0$ & $0$ & $0$ & $-1/3$ & $1/3$ & $1/3$ & $-1/3$ & $0$ & $0$ & $1/3$\\
\end{tabular}
\vspace{1mm}
\caption{The charge assignment of fields in the low-energy effective theory.}
\label{tab:charge2}
\end{table*}

At low-energies, the theory shows ``smooth" confinement or s-confinement \cite{Csaki:1996sm,Csaki:1996zb}
and is described by $N_F^2$ mesons and $N_F$ pairs of (anti-)baryons,
\beq
\begin{split}
&\mathcal{M}'^{~\beta}_{\alpha} \equiv \bar{Q}^{\beta n} Q_{\alpha n} \, , \\[1ex]
&B'^\alpha \equiv \epsilon^{\alpha_1,\alpha_2,\cdots,  \alpha_N, \alpha}
Q_{\alpha_1 n_1}\cdots Q_{\alpha_N n_N}\epsilon^{n_1,\cdots, n_N} \, , \\[1ex]
&\bar{B}'_\alpha \equiv \epsilon_{\alpha_1,\alpha_2,\cdots,  \alpha_N, \alpha}
\bar{Q}^{\alpha_1 n_1}\cdots \bar{Q}^{\alpha_N n_N}\epsilon_{n_1,\cdots, n_N} \, .
\end{split}
\eeq
Here, $\epsilon$ is a totally anti-symmetric tensor.
The field with a prime $``~'~"$ is not canonically normalized.
With these composite fields, the effective superpotential is given by
\cite{Seiberg:1994bz,Seiberg:1994pq}
\beq
W_{\rm eff} &\supset& \frac{1}{\Lambda^{9}} \lmk B'^\alpha\mathcal{M}'^{~\beta}_{\alpha}\bar{B}'_\beta - \det\mathcal{M}'\rmk\nonumber\\
    &+& \frac{\widetilde{m}}{3} \mathcal{M}'^{~p}_{p} 
+ \widetilde{M}^{~i}_{j} \mathcal{M}'^{~j}_{i}+\widetilde{\iota}_j^{~i}\Psi_i^{~p}{\mathcal{M}'}_p^{~j}\nonumber\\
    &+& \frac{\widetilde{\rho}}{3\mu}\mathcal{M}'^{~p}_q\mathcal{M}'^{~q}_p
    +\frac{1}{\mu}\widetilde{\lambda}_{jl}^{~ik}{\mathcal{M}'_{i}}^{j}{\mathcal{M}'_{k}}^{l}
\, \ .
\label{Weff}
\eeq
Canonically normalizing the composite fields,
$\mathcal{M}' \rightarrow \mathcal{M}$, $B' \rightarrow B$, $\bar{B}' \rightarrow \bar{B}$,
we now define
\beq
&&\frac{1}{3}\mathcal{M}_{p}^{~p} \equiv \Phi \, ,~~ \mathcal{M}_p^{~q} -\Phi\delta_p^{~q} \equiv \tilde{\Phi}_p^{~q} \, , ~~ \mathcal{M}^{~j}_{i} \equiv \Sigma_{i}^{~j} \, ,\nonumber\\[1ex]
&&\mathcal{M}_p^{~j} \equiv \Xi_p^{~j} \, , ~~
\mathcal{M}_{i}^{~q} \equiv D^{~q}_{i} \, , \\[1ex]
&&B^\alpha \equiv
\lmk
\begin{array}{c}
\bar{T}^p\\
F^i\\
\end{array}
\rmk,~~~
\bar{B}_\alpha \equiv
\lmk
\begin{array}{c}
T_p\\
\bar{F}_i\\
\end{array}
\rmk ,
\eeq
to reproduce the superpotential of Eq.~\eqref{HS} for the Hiller-Schmaltz mechanism.
In particular, $D_i^{~q}$ are identified as the right-handed down-type quarks.
It determines the hypercharges of the constituent fields,
\beq
Y_{Q_p} = -\frac{2}{15} \, , \qquad Y_{Q_i} = \frac{1}{5} \, .
\eeq
The charge assignment of fields in the low-energy effective theory is summarized in Table \ref{tab:charge2}.
Thus, one can see that the first term in \EQ{Weff} includes the same superpotential with the original model of \EQ{HS},
\beq
W_{\rm eff} &\supset& r^{~i}_{j}D^{~p}_iF^jT_p + y_\Phi \Phi T_p\bar{T}^p + y_{\Sigma}\Sigma^{~j}_{i}F^i\bar{F}_j\nonumber\\
&+& \tilde{y}_{\Phi}\tilde{\Phi}_p^{~q}T_q\bar{T}^p + y_\Xi\Xi_p^{~j}\bar{F}_j\bar{T}^p -\frac{(4\pi)^{4}c}{\Lambda^{3}}\det\mathcal{M}\nonumber\\
&+& \frac{\Lambda}{4\pi} \lmk m\Phi + M_j^{~i}\Sigma_i^{~j} \rmk +\frac{\Lambda}{4\pi}\iota_j^{~i}\Psi_i^{~p}\Xi_p^{~j}\nonumber\\
&+& \frac{\Lambda^2}{\mu}\lmk \rho\Phi^2 +\frac{\rho}{3} \Tr\tilde{\Phi}^2 + \lambda_{jl}^{~ik}\Sigma_i^{~j}\Sigma_k^{~l}\rmk ,
\label{Wfull}
\eeq
where $c$ is given by an $\mathcal{O}(1)$ number and
we have defined ${\iota}, {\rho}, {\lambda}, {m}, M$ to absorb $\mathcal{O}(1)$ factors in front of the corresponding parameters.
The coefficient $\lambda$ respects a symmetry $\lambda_{jl}^{~ik}=\lambda_{lj}^{~ki}$.
By using NDA, the other coefficients are estimated as
\beq
r_{j}^{~i} \sim y_\Phi \sim \tilde{y}_\Phi \sim y_\Sigma \sim y_\Xi \sim 4\pi \, .
\eeq
Since these Yukawa couplings are large, as discussed in the introduction,
if the matrix $\Sigma$ acquires a VEV, which is complex and not close to the identity matrix,
and also $y_\Sigma|\langle\Sigma\rangle| \sim y_\Phi\langle\Phi\rangle\sim M_{\rm CP}$,
the correct CKM phase is induced.

As the right-handed down-type quarks are composite in our model,
their masses are generated through higher-dimensional terms,
\beq
W &\supset& \frac{q^{~j}_p{\widetilde{Y}_{Dj}}^{~~i}H_d(\bar{Q}^{pn}Q_{in})}{\mu'} \label{UVyukawa} \\
\rightarrow W_{\rm eff} &\supset&
\frac{1}{4\pi}\frac{\Lambda}{\mu'} q_p^{~j}{{Y}_{Dj}}^{i} D_i^{~p}H_d \, ,
\label{Yukawa}
\eeq
where $q$ and $H_d$ denote the left-handed quark doublets and the down-type Higgs doublet, respectively,
$\mu'$ is a mass parameter, $\widetilde{Y}_D$ is a dimensionless coupling and $Y_D$ is defined
to absorb an $\mathcal{O}(1)$ factor in front of the corresponding coupling.
To achieve the correct bottom quark mass requires 
\beq
\Lambda \lesssim \mu'\lesssim 10^{2}\times \frac{\Lambda}{4\pi} \, . \label{muprimerelation}
\eeq
A possible UV completion of Eq.~\eqref{UVyukawa} satisfying the relation \eqref{muprimerelation}
is discussed in Appendix~\ref{app:UV}.

At the renormalizable level, it is possible to include a superpotential term,
$W\supset q_p^{~j}{{Y}_{\Psi j}}^{i} \Psi_i^{~p}H_d$, with $Y_\Psi$ a real Yukawa coupling constant for $\Psi$. 
Including this term, the mass terms for the down-type quarks after the electroweak symmetry breaking are given by
\beq
-\Lag_M = (D~\Psi) M_{D\Psi} \lmk
\begin{array}{c}
D_L \\
\Xi \\
\end{array}
\rmk ,
\eeq
with the $6\times6$ mass matrix,
\beq
M_{D\Psi} =
\lmk
\begin{array}{cc}
\frac{1}{4\pi} \frac{\Lambda}{\mu'}Y_D v_d & 0  \\
Y_\Psi v_d & \frac{\Lambda}{4\pi}\iota \\
\end{array}
\rmk,
\eeq
where $D_L$ denotes the down-type component of the left-handed quark and $v_d$ is the VEV of $H_d$.
This matrix structure is reminiscent of the Nelson-Barr model \cite{Nelson:1983zb,Barr:1984qx},
although there is an essential difference: all the components of our mass matrix are completely real.
In the basis which diagonalizes the mass matrix, we can decompose $D,D_L,\Psi,\Xi$ into lighter $(d,d_L)$ and heavier states $(\psi,\xi)$.
The mass of $(\psi,\xi)$ is at the order of $\Lambda$.
Since the mixing angle is suppressed by $v_d/\Lambda$ that is naturally much smaller than the unity, we identify $D,D_L\sim d, d_L$ and $\Psi,\Xi\sim\psi,\xi$ in the following.

\section{Vacuum structure
\label{sec:vacuum}}

We now investigate the vacuum structure of the model with the superpotential (\ref{Wfull})
and demonstrate that it accommodates spontaneous CP violation to realize the observed CKM phase.
The stabilization of the desired vacuum solution is described, while its details are given in Appendix \ref{app:stabilization}.

Since all the $SU(3)_c$ color charged fields are massive,
their VEVs are set to zero.
In fact, the $SU(3)_c$ adjoint $\widetilde{\Phi}$ acquires a mass from the term $\Tr\widetilde{\Phi}^2$, and $\langle\widetilde{\Phi}\rangle=0$.
As we have discussed in the previous section,
$\Psi$ and $\Xi$ obtain a large mass term, so that $\langle\Xi\rangle=\langle\Psi\rangle=0$. 
The $\mathcal{F}$-term conditions, $\mathcal{M}_\alpha^{~\beta} B^\alpha=\mathcal{M}_\alpha^{~\beta}\bar{B}_{\beta}=0$,
guarantee $\langle T \rangle = \langle \bar{T} \rangle = \langle F \rangle = \langle \bar{F} \rangle=0$.
Thus, the vacuum structure is determined by the superpotential,
\beq
W_{\rm eff} (\Phi, \Sigma) &=& -\frac{(4\pi)^{4}c}{\Lambda^{3}}\Phi^3\det\Sigma\nonumber\\
&+& \frac{\Lambda}{4\pi} \lmk m\Phi + M_j^{~i}\Sigma_i^{~j} \rmk\nonumber\\
&+& \frac{\Lambda^2}{\mu}\lmk\rho\Phi^2 + \lambda_{jl}^{~ik}\Sigma_i^{~j}\Sigma_k^{~l}\rmk.
\label{effsup}
\eeq
The $\mathcal{F}$-term conditions for $\Phi, \Sigma$ are then given by
\beq
&&\frac{(4\pi)^5c}{\Lambda^{4}} \Phi^2\det\Sigma = \frac{1}{3}m + \frac{8\pi\rho}{3}\frac{\Lambda}{\mu}\Phi \, ,\label{min1}\\[1ex]
&&\frac{(4\pi)^5c}{\Lambda^{4}} \Phi^3 \tilde{\Sigma}_i^{~j} = 
M_i^{~j} + 
8\pi\lambda_{ik}^{~jl}\frac{\Lambda}{\mu}\Sigma_l^{~k} \, .
\label{min2}
\eeq
Here, $\tilde{\Sigma}_i^{~j}$ denotes the $(i,j)$-component of adjugate matrix of $\Sigma$.\footnote{Note that transpose is taken in the definition of adjugate matrix, i.e.,  the cofactor, ${\rm cof}(\Sigma_j^{~i})$, of $(j,i)$-component is related as $\tilde{\Sigma}_i^{~j}\equiv{\rm cof}(\Sigma_j^{~i})$.}

\subsection{Vacuum moduli space}

For the Hiller-Schmaltz mechanism to work well in our model,
we require a complex VEV of $\Sigma$ and a real VEV of $\Phi$ with similar magnitudes.
To see how to obtain such a solution,
let us first take a specific form of $\lambda_{ik}^{~jl}=\lambda_0\delta_i^l\delta_k^j$ and $M_i^{~j} = M\delta_i^j$,
with $\lambda_0$ and $M$ real parameters, which clarifies the magnitudes of VEVs and the structure of vacua.
Here, we also assume $\rho>0,\lambda_0>0$.\footnote{We can take a different set of signs, i.e. $\rho\cdot\lambda_0<0$, so that the proper solutions can be obtained. However, as shown in \EQ{eq:CPscale}, we need $\rho\cdot\lambda_0>0$ in the simplified case of $m=M_i^j=0$.} 
Note that $\det \langle \Sigma \rangle \in \mathbb{R}$ is satisfied because of \EQ{min1} with $\langle\Phi\rangle\in\mathbb{R}$. 
Diagonal components of Eq.~(\ref{min2}) are then reduced to 
\beq
\frac{(4\pi)^5c}{\Lambda^{4}} \Phi^3 \tilde{\Sigma}_i^{~i} = 
M + 
8\pi\lambda_0\frac{\Lambda}{\mu}\Sigma_i^{~i} \, ,
\label{diageq}
\eeq
where the index $i$ is not summed over.
Using the equations for $i=1 , 2$, we obtain 
\beq
\Sigma_3^{~3} = -\frac{2\lambda_0}{(4\pi)^4c}\frac{\Lambda^5}{\mu}\Phi^{-3} \, .
\eeq
Likewise, we can see that $\Sigma_2^{~2}=\Sigma_3^{~3}$.
Substituting these solutions into Eq.~\eqref{diageq}, the $(1,1)$-component of $\Sigma$ is given by
\beq
\Sigma_1^{~1} = -\frac{M}{8\pi\lambda_0}\frac{\mu}{\Lambda}+\frac{2\lambda_0}{(4\pi)^4c} \frac{\Lambda^5}{\mu}\Phi^{-3} \, .
\eeq
By substituting these solutions of $\Sigma_i^{~i}$ and $\Sigma_i^{~j} = 0$ for $i \neq j$ to \EQ{min1},
we find a VEV of $\Phi$.
This set of VEVs obviously produces no complex phase to break CP symmetry.
However, Eqs.~(\ref{min1}), (\ref{min2}) actually accommodate degrees of freedom to transform
this real-valued diagonal solution for $\Sigma$ by a three-dimensional general linear group ${\rm GL}(3,\mathbb{C})$,
$\vev\Sigma \rightarrow U^{-1}\vev\Sigma U$ where $U \in {\rm GL}(3,\mathbb{C})$,
so that $\Sigma^\dagger\Sigma$ is complex-valued.
In other words, there exists a vacuum moduli space in our model with
the specific choice of $\lambda_{ik}^{~jl}=\lambda_0\delta_i^l\delta_k^j$ and $M_i^{~j} = M\delta_i^j$.

To estimate the magnitudes of VEVs,
we further take $m=M_i^{~j}=0$ for simplicity, substitute the solutions to \EQ{min1} and find
\beq
\langle\Phi\rangle &=& \lmk\frac{\lambda_0\Lambda}{\mu}\rmk^{\frac{1}{4}} \lmk\frac{3\lambda_0}{2\rho}\rmk^{\frac{1}{8}}  \frac{\Lambda}{4\pi} \, ,\\[1ex]
|\langle\Sigma\rangle| &=& \lmk\frac{\lambda_0\Lambda}{\mu}\rmk^{\frac{1}{4}} \lmk\frac{3\lambda_0}{2\rho}\rmk^{\frac{3}{8}}  \frac{\Lambda}{4\pi} \, ,
\label{eq:CPscale}
\eeq
where the scale of $\Sigma$ is defined as the eigenvalue.
We can see that the values of $\Phi$ and $\Sigma$ have the same magnitude.
For $\rho\sim\lambda_0 \sim 1$, we obtain $M_{\rm CP} \sim y_\Phi\langle\Phi\rangle \sim y_\Sigma |\langle\Sigma\rangle| \sim (\Lambda/\mu)^{1/4}\Lambda$,
which is below the scale of confinement.

\subsection{Moduli stabilization}

Let us now consider the stabilization of the continuous moduli space around a desired minimum of $\vev\Sigma$.
Here, we take a general form of $M_i^{~j}$ and $\lambda_{ik}^{~jl}$.
Since the $\mathcal{F}$-term condition for $\Sigma$, ${\partial W_{\rm eff}}/{\partial \Sigma} = 0$ at $\Sigma = \vev\Sigma$, is satisfied, we can rewrite the superpotential \eqref{effsup} as
\begin{eqnarray}
\label{Weffmoduli}
    W_{\rm eff} &=& W_{\rm eff} - \left.\dfrac{\partial W_{\rm eff}}{\partial \Sigma}\right|_{ \vev\Sigma} (\Sigma - \vev\Sigma) \nonumber \\[1ex]
    &=& W_0 + \delta W + {\rm const.} \, , 
\end{eqnarray}
where
\begin{eqnarray}
    W_0 &=& -\dfrac{(4\pi)^4c}{\Lambda^3}\Phi^3 
    \left(\det\Sigma - \tilde{\vev\Sigma}\delta\Sigma \right)\nonumber ,\label{eq:W0}\\[1ex]
    \delta W &=& \dfrac{\Lambda^2}{\mu} \lambda_{jl}^{~ik} \delta\Sigma_i^{~j} \delta\Sigma_k^{~l}
    \, . \label{eq:delta W}
\end{eqnarray}
Here, $\delta\Sigma \equiv \Sigma - \vev\Sigma$, and
in this parameterization, we can retrieve the mass parameter $M_i^{~j}$ in terms of a VEV $\vev\Sigma$ as
\begin{equation}
    M=\frac{(4\pi)^4c}{\Lambda^3}\Phi^3 \tilde{\vev{\Sigma}} - 2 \frac{\Lambda^2}{\mu} \lambda \vev\Sigma.\label{eq:Mass}
\end{equation}
Note that all couplings in the superpotential, including $\lambda_{jl}^{~ik}$ and $M_i^{~j}$, are real due to CP invariance.

The $W_0$ part of the superpotential \eqref{Weffmoduli} only affects the eigenvalues of $\vev\Sigma$, and is immune to
the transformation of ${\rm GL}(3,\mathbb{C})$ around $\vev\Sigma$.
This is obvious for the $\det\Sigma$ term.
For the second $\tilde{\vev\Sigma}\delta\Sigma$ term, an infinitesimal ${\rm GL}(3,\mathbb{C})$ transformation,
$\Sigma=U^{-1}\vev\Sigma U$, gives
$\tilde{\vev\Sigma}\delta\Sigma \approx \det(U^{-1}\vev\Sigma U) - \det(\vev\Sigma) = 0$.
Such ${\rm GL}(3,\mathbb{C})$ degrees of freedom are used to obtain a desired vacuum where
$\vev\Sigma^\dagger\vev\Sigma$ is complex-valued.
The $\delta W$ part of the superpotential is responsible for the stabilization of the moduli space of ${\rm GL}(3,\mathbb{C})$ degrees of freedom.
In Appendix \ref{app:stabilization}, 
we will show that if we choose a proper $ \lambda_{jl}^{~ik}$, the moduli space of $\vev\Sigma$ can be completely stabilized
at a desired $\vev\Sigma$ to spontaneously break CP symmetry.

\section{Corrections to $\bar{\theta}$
\label{sec:constraint}}

Although the non-renormalization theorem forbids radiative corrections to $\bar{\theta}$,
it is not guaranteed after SUSY breaking.
Furthermore, some Planck-suppressed operators may destroy our solution to the strong CP problem.
Here, we discuss these issues and find a condition on the scale of spontaneous CP violation in our model.

\subsection{SUSY breaking}

We assume that the mediation scale of SUSY breaking is lower than the scale of spontaneous CP violation.
A hidden SUSY breaking sector is CP-symmetric so that soft SUSY breaking parameters in the visible sector do not
contain CP-violating complex phases at the mediation scale. 
This is achieved by a low-scale gauge-mediated SUSY breaking model \cite{Giudice:1998bp}.
Then, at the scale of $M_{\rm CP}$, $\Psi$ and $\Xi$ are integrated out supersymmetrically,
and the effective theory is reduced to the original Hiller-Schmaltz model (\ref{HS}).
Following the discussion of ref.~\cite{Hiller:2002um}, with soft SUSY breaking parameters,
radiative corrections can induce CP-violating phases to the gluino and quark masses,
which contribute to $\bar{\theta}$.
To avoid such dangerous corrections requires a hierarchy between the CP violation scale
and the mediation scale of SUSY breaking,
\beq
M_{\rm SUSY}\lesssim 10^{-3}M_{\rm CP} \, .
\label{hierarchy}
\eeq
For $M_{\rm SUSY} \gtrsim 10^5 \GeV$, we obtain a condition, $M_{\rm CP} \gtrsim 10^8\GeV$.

\subsection{Planck-suppressed operators}

We now consider all possible Planck-suppressed operators respecting CP and gauge symmetries of the model
and check if they do not destroy the solution to the strong CP problem. 
One of the most relevant operators is
\beq
\lmk\frac{\bar{Q}^{jn}Q_{in}}{\Mpl^2}\rmk W_\alpha W^\alpha,
\eeq
where $W_\alpha$ denotes the field strength chiral superfield for $SU(3)_c$.
After the confinement, this operator leads to 
\beq
\lmk\frac{\Lambda^2}{16\pi^2\Mpl^2}\rmk \lmk\frac{\Sigma_i^{~j}}{\Lambda/4\pi}\rmk W_\alpha W^\alpha.
\eeq
With the VEV of $\Sigma$, it directly contributes to the $\theta$ term.
Assuming $\mu \sim \Lambda$, the size of the contribution is estimated as $(\Lambda/4\pi\Mpl)^2$,
and the current upper limit on $\bar{\theta}$ imposes $\Lambda\lesssim 10^{14}\GeV$.
Note that our composite model gets an extra suppression factor of $(\Lambda/\Mpl)$
compared to the case of the original Hiller-Schmaltz model
\cite{Hiller:2001qg, Hiller:2002um}.

Some Planck-suppressed operators can contribute to the down-type quark mass matrix
and generate corrections to $\bar{\theta}$.
They include
\beq
&&W\supset \widetilde{\bar{\iota}}_j^{~i}\frac{(\bar{Q}^{p}Q_{i}) (\bar{Q}^{j}Q_p)}{\Mpl} 
+  \widetilde{\varepsilon}_{jl}^{~ik} \Psi_i^{~p} \frac{(\bar{Q}^{l}Q_{k})(\bar{Q}^{j}Q_{p})}{\Mpl^2}\nonumber\\
&\rightarrow& W_{\rm eff} \supset
\frac{\Lambda^2}{\Mpl}\bar{\iota}_j^{~i}D_i^{~p} \Xi_p^{~j}+ \lmk\frac{\Lambda}{\Mpl}\rmk^2 \varepsilon_{jl}^{~ik}\Sigma_k^{~l}\Psi_i^{~p}\Xi_p^{~j} \, ,\nonumber\\
\label{extra}
\eeq
where $\widetilde{\bar{\iota}}, \widetilde{\varepsilon}$ are dimensionless coupling constants, and $\bar{\iota}, \varepsilon$ are defined to absorb $\mathcal{O}(1)$ factors.
The mass matrix for $(D,\Psi), (D_L, \Xi)$ is then given by
\beq
M_{D\Psi}\simeq
\lmk
\begin{array}{cc}
\frac{1}{4\pi} \frac{\Lambda}{\mu'}Y_D v_d & \frac{\Lambda^2}{\Mpl}\bar{\iota} \\
Y_\Psi v_d & \frac{\Lambda}{4\pi} \lmk \iota + 4\pi\frac{\Lambda}{\Mpl^2}\varepsilon\langle\Sigma\rangle \rmk \\
\end{array}
\rmk.
\eeq
Since the only source for CP violation lies in the right-lower component, the contribution of $\arg\det M_{D\Psi}$
is estimated as $\delta\bar{\theta} \sim (\Lambda/\Mpl)^2$, which requires $\Lambda\lesssim 10^{13}\GeV$.
In summary, $10^8\GeV \lesssim M_{\rm CP} \lesssim 10^{12}\GeV$ can give the viable solution
to the strong CP problem in our model.

\section{Discussions
\label{sec:discussion}}

We have pursued a solution to the strong CP problem by exploring a composite model of spontaneous CP violation
based on a new SUSY QCD.
The complex VEVs of the mesons break CP symmetry spontaneously.
The scale of spontaneous CP violation is then hierarchically lower than the Planck scale.
The Hiller-Schmaltz mechanism provides the CKM phase without generating the strong CP phase $\bar{\theta}$.
In our model, the right-handed down-type quarks are composite,
enhancing their couplings to CP breaking fields, which is essential to realize the observed CKM phase.
We have identified the range of the CP violation scale where the strong CP problem is correctly resolved.
Since SUSY can also provide a solution to the electroweak naturalness problem and a candidate of the cosmological dark matter
(the gravitino in our case),
the present scenario may give a serious candidate of physics beyond the SM.

In the early Universe, if CP symmetry is spontaneously broken after inflation, it leads to the production of domain walls. 
Such domain walls dominate the Universe soon after the generation, and spoils the standard cosmology.
The situation even gets worse when CP symmetry is gauged, which is demanded for an exact symmetry
\cite{Harlow:2018jwu}.
In that case, the domain wall is necessarily stable
\cite{McNamara:2022lrw,Asadi:2022vys}.
Thus, CP symmetry must be broken during inflation and no longer restored after inflation. 
The maximal temperature in the Universe after inflation is given by
\beq
T_{\rm max} \sim \lmk\frac{\Gamma_{\rm inf}}{H_{\rm inf}}\rho_{\rm inf}\rmk^{\frac{1}{4}}\sim (T_{\rm RH}^2H_{\rm inf}\Mpl)^{\frac{1}{4}} \, ,
\eeq
where $\Gamma_{\rm inf}~(\simeq T_{\rm RH}^2/\Mpl)$ and $\rho_{\rm inf}~(\simeq H_{\rm inf}^2\Mpl^2)$ respectively denote
the inflaton decay rate and the energy density during inflation,
$H_{\rm inf}$ is the Hubble parameter during inflation and $T_{\rm RH}$ is the reheating temperature.
The mass of $\Sigma$, defined as $m_\Sigma$, is then required to be larger than the maximal temperature,
\beq
m_{\Sigma} \sim \lmk\frac{M_{\rm CP}}{\mu}\rmk^{\frac{3}{5}} M_{\rm CP} \gtrsim T_{\rm max} \, ,
\eeq
or the reheating temperature is bounded as
\beq
T_{\rm RH} &\lesssim& 1\times10^{7}\GeV \lmk \frac{M_{\rm CP}}{10^{12}\GeV}\rmk^{\frac{13}{5}}\nonumber\\ 
&\times& \lmk\frac{\mu}{10^{13}\GeV}\rmk^{-\frac{6}{5}}\lmk\frac{H_{\rm inf}}{10^{13}\GeV}\rmk^{-\frac{1}{2}}.
\eeq
For a low messenger mass scale of gauge mediation,
this bound is looser than that from the overproduction of light gravitinos
\cite{Moroi:1993mb}.
In other words, with an appropriate reheating temperature, the produced gravitino can give the correct abundance of
dark matter without encountering the domain wall problem. 
For the generation of baryon asymmetry, thermal leptogenesis
\cite{Fukugita:1986hr} does not work, while Affleck-Dine baryogenesis
\cite{Affleck:1984fy} seems to be promising.

\section*{Acknowledgments}
We thank Masaki Yamada and Yufei Zhang for useful discussions.

\appendix

\section{Details of moduli stabilization
\label{app:stabilization}}
The moduli space $\mathfrak M$ is the manifold for $\Sigma$
connected to a desired VEV $\vev\Sigma$
by the transformation of ${\rm GL}(3,\mathbb{C})$,
i.e. $\mathfrak M = \{\Sigma\,|\Sigma=U^{-1}\vev\Sigma U\}$ for all $U\in {\rm GL}(n_\Sigma,\mathbb{C})$
where $n_\Sigma =3$.
Since similarity transformations do not change the eigenvalues of $\vev\Sigma$,
$n_\Sigma$ degrees of freedom in $\Sigma$ are already fixed by $W_0$ in Eq.~\eqref{eq:delta W}.
Then, the number of remaining degrees of freedom is $n_\Sigma^2-n_\Sigma$,
which must be stabilized by $\delta W$.
In this appendix, we will clarify the structure of the moduli space $\mathfrak M$ and perform its stabilization around a desired point $\vev\Sigma$ to break CP symmetry spontaneously.

Note that the $n_\Sigma \times n_\Sigma$ complex matrix $\Sigma$ lives in a $n_\Sigma^2$-dimensional linear space $\mathcal V$,
and any $\Sigma$ can be expressed by a linear combination of $n_\Sigma^2$ linearly independent bases.
Here, we choose the so-called Cartan-Weyl basis $\{\sigma_a\}$ for $\mathcal V$.
For $n_\Sigma=3$, the index $a$ runs over $0,1,\cdots , 8$.
Such $\{\sigma_a\}$ are composed of three Cartan subalgebra elements $H_{0,1,2}$,\footnote{The Cartan subalgebra of ${\rm GL}(3,\mathbb C)$ is spanned by the generator of its center $H_0$ and the Cartan subalgebra of $SU(3)$.}
and six ladder operators $E_{\pm t, \pm u, \pm v}$.
Up to an overall similarity transformation, $\{\sigma_a\}$ can be written as
\begin{eqnarray}
    \sigma_0 = H_0 = \frac{1}{\sqrt 3}{\mathbf 1} \,
,&\label{eq:Sigma basis} \nonumber \\[1ex]
\sigma_1 = H_1 = \frac1{\sqrt 2}\left(
\begin{array}{ccc}
 1 & 0 & 0 \\
 0 & -1 & 0 \\
 0 & 0 & 0 \\
\end{array}
\right),\,&
\sigma_2 = H_2 = \frac1{\sqrt 6} \left(
\begin{array}{ccc}
 1 & 0 & 0 \\
 0 & 1 & 0 \\
 0 & 0 & -2 \\
\end{array}
\right),\nonumber\\[1ex]
\sigma_3 = E_{+t} = \left(
\begin{array}{ccc}
 0 & 1 & 0 \\
 0 & 0 & 0 \\
 0 & 0 & 0 \\
\end{array}
\right),\, &
\sigma_4 = E_{-t} =\left(
    \begin{array}{ccc}
     0 & 0 & 0 \\
     1 & 0 & 0 \\
     0 & 0 & 0 \\
    \end{array}
    \right),\nonumber\\[1ex]
\sigma_5 = E_{+u} = \left(
    \begin{array}{ccc}
        0 & 0 & 0 \\
        0 & 0 & 1 \\
        0 & 0 & 0 \\
    \end{array}
    \right),\, &
\sigma_6 = E_{-u} =\left(
        \begin{array}{ccc}
            0 & 0 & 0 \\
            0 & 0 & 0 \\
            0 & 1 & 0 \\
        \end{array}
    \right),\nonumber\\[1ex]
\sigma_7 = E_{+v} = \left(
    \begin{array}{ccc}
        0 & 0 & 1 \\
        0 & 0 & 0 \\
        0 & 0 & 0 \\
    \end{array}
    \right),\, &
\sigma_8 = E_{-v} =\left(
        \begin{array}{ccc}
            0 & 0 & 0 \\
            0 & 0 & 0 \\
            1 & 0 & 0 \\
        \end{array}
    \right). \nonumber \\
    {}
\end{eqnarray}
This basis has an inner product defined by
\begin{equation}
    \langle \sigma_a, \sigma_b \rangle =  \Tr(\sigma^a \sigma_b)  =\delta^a_{b} \, ,
\end{equation}
where $\sigma^a = g^{ab}\sigma_b$ and $g^{ab}$ is the inverse of the Cartan-Killing metric, $g_{ab} = \Tr(\sigma_a \sigma_b)$.
In the current basis, it is
\begin{equation}
    g_{ab} = \left(
    \begin{array}{ccc|ccc|ccc}
     1 &      &      &      &      &      &      &      &      \\
          & 1 &      &      &      &      &      &      &      \\
          &      & 1 &      &      &      &      &      &      \\\hline
          &      &      &      &      &      & 1 &      &      \\
          &      &      &      &      &      &      & 1 &      \\
          &      &      &      &      &      &      &      & 1 \\\hline
          &      &      & 1 &      &      &      &      &      \\
          &      &      &      & 1 &      &      &      &      \\
          &      &      &      &      & 1 &      &      &      \\
    \end{array}
    \right) ,
\end{equation}
for $a,b=0,1,\cdots , 8$.
With the above inner product, we can decompose any $\Sigma$ like a vector,
\begin{equation}
    \Sigma_j^{~i} = \Sigma^a (\sigma_a)_j^{~i} \, , \label{sigmaa}
\end{equation}
with $i,j$ running over $1,\cdots , n_\Sigma=3$.

\begin{figure}
\label{modulispacefigure}
\centering
\includegraphics[width=0.27\textwidth]{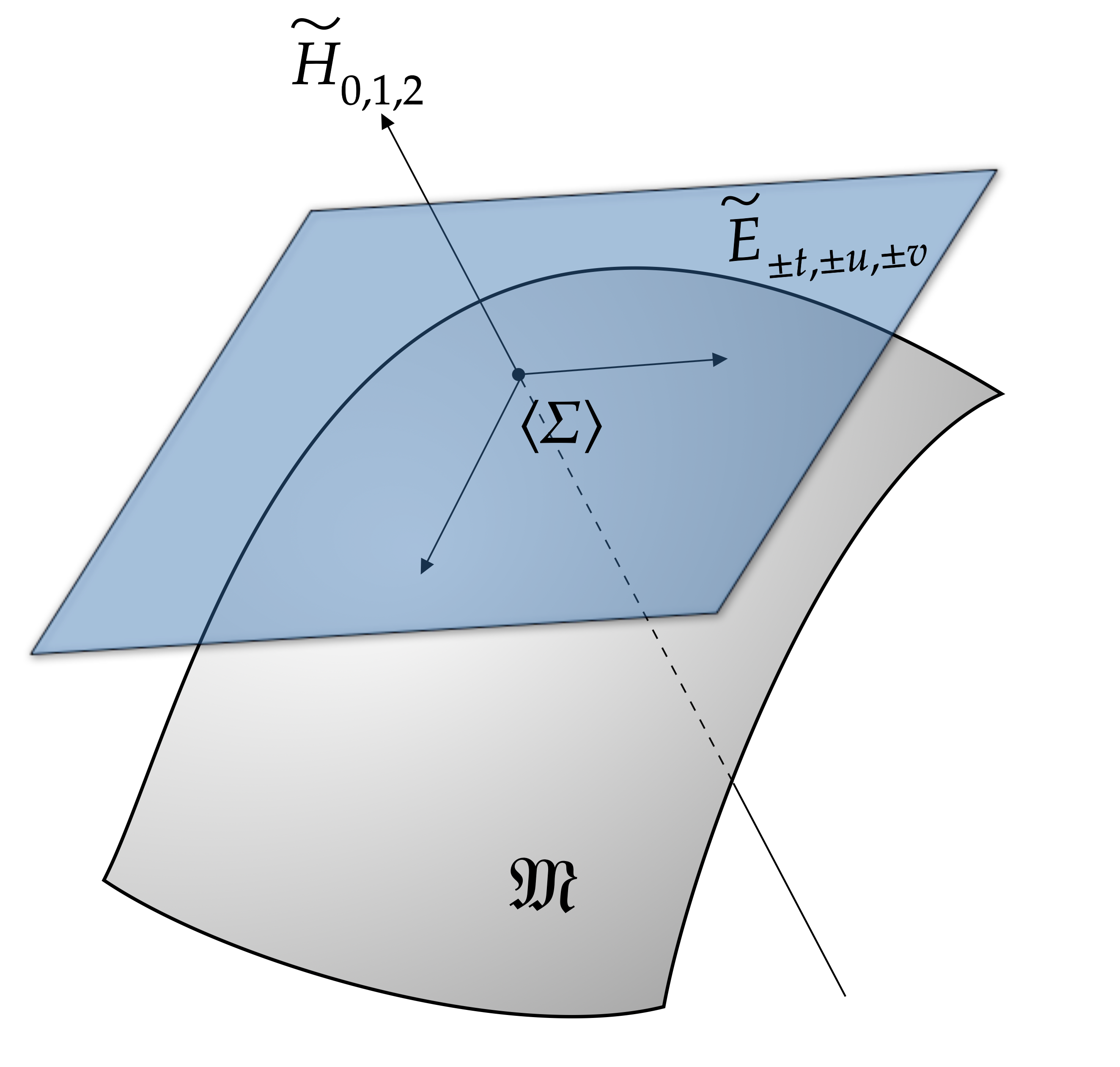}
\caption{The tangent and normal space of $\mathfrak M$ at $\vev\Sigma$. The basis of the tangent space is given by $\tilde E_{\pm t, \pm u, \pm v}$, and the normal space is spanned by $\tilde H_{0,1,2}$.}
\label{Fig-tangent}
\end{figure}

For $\Sigma \in \mathfrak M$, not all $\Sigma^a$ in Eq.~\eqref{sigmaa} are independent,
and they are subject to the constraint, $\Sigma = U^{-1}\vev\Sigma U$.
We can build an intrinsic coordinate system locally for $\mathfrak M$ around $\vev\Sigma$,
by parameterizing $U$ as
\begin{equation}
    U(\theta) = e^{i\theta^a \sigma_a} 
    \approx \mathbf 1 + i \theta^a \sigma_a \, .
\end{equation}
Here, $\theta^a$ ($a=0,1,\cdots , 8$) are complex infinitesimal parameters.
A group element $U(\theta)$ acts on $\Sigma$ in the adjoint representation,
\begin{eqnarray}
    \Sigma(\theta)_j^{~i} &=& U^{-1}(\theta)_j^{~j'}\langle\Sigma\rangle_{j'}^{~i'} U(\theta)_{i'}^{~i}\nonumber \\
    &=& \langle\Sigma\rangle_{j}^{~i} 
    - i \theta^a \left[ \sigma_a, \langle\Sigma\rangle \right]_{j}^{~i}
    + \frac12  \theta^a \theta^b \left[\sigma_a, \left[ \sigma_b, \langle\Sigma\rangle \right]\right]_{j}^{~i} +\cdots \nonumber\\
    &=& \langle\Sigma\rangle_{j}^{~i}  + i \theta^a D_a \Sigma_{j}^{~i} + \frac12 \theta^a \theta^b D_a D_b \Sigma_{j}^{~i} +\cdots  ,\label{eq:SL transformation}
\end{eqnarray}
where the Baker-Campbell-Hausdorff formula is used on the second line.
We can see that $\theta^a$s give a local coordinate for $\mathfrak M$ around $\vev\Sigma$,
and its Lie derivatives are given by the commutators with $\vev\Sigma$, 
i.e. $D_a = \partial/\partial \theta^a = [\cdot,\,\sigma_a]$.
It is not surprising that there exist only $n_\Sigma^2-n_\Sigma=6$ non-vanishing $D_a$ when acting on $\Sigma$,
because they correspond to the basis of the tangent space for the moduli space $\mathfrak M$.
To see this explicitly, we first assume the VEV is diagonal, so that $\vev\Sigma_{\rm diag}$ only has three non-vanishing components corresponding to three diagonal bases $H_0, H_1, H_2$,
\begin{equation}
    \vev\Sigma_{\rm diag} = \vev\Sigma_{\rm diag}^0 H_0 + \vev\Sigma_{\rm diag}^1 H_1 + \vev\Sigma_{\rm diag}^2 H_2 \, .\label{eq:diag Sigma diag}
\end{equation}
These diagonal bases form a Cartan subalgebra,
and they commute with each other, $[H_i, H_j]=0$ for $i,j=0,1,2$.
Then three out of nine Lie derivatives $D_0, D_1, D_2$ become zero.
This feature is preserved under an overall similarity transformation $U_0$,
which makes $\vev\Sigma$ a non-diagonalized form:
\begin{eqnarray}
    \vev\Sigma &=& U^{-1}_0\vev\Sigma_{\rm diag} U_0\nonumber\\
    &=& U^{-1}_0\left(\vev\Sigma_{\rm diag}^i H_i\right) U_0\nonumber\\
    &=& \vev\Sigma_{\rm diag}^i \left(U^{-1}_0 H_i U_0 \right)=\vev\Sigma_{\rm diag}^i \tilde{H}_i \, , \label{eq:diag Sigma}
\end{eqnarray}
with a new basis $\{\tilde \sigma_a\} = \{\tilde H, \tilde E\}$
where $\tilde \sigma_a = U_0^{-1}\sigma_a U_0$.
The Lie algebra and the inner product are invariant under an overall similarity transformation, and
$\{\tilde H_i\}$ still form a Cartan subalgebra of ${\rm GL}(3,\mathbb C)$.
Therefore, for an arbitrary $\vev\Sigma$,
the first three Lie derivatives $D_0, D_1, D_2$ corresponding to the Cartan subalgebra vanish.
Indeed, these Lie derivatives $D_0, D_1, D_2$ are related to three $\vev\Sigma$ eigenvalue degrees of freedom,
which are fixed by the superpotential $W_0$.
Geometrically, $D_0, D_1, D_2$ are normal directions of $\mathfrak M$ at $\vev\Sigma$.
See Figure~\ref{Fig-tangent}.

Except for these normal directions $D_0, D_1, D_2$, the remaining six Lie derivatives $D_{\pm t, \pm u, \pm v}$ are tangential to $\mathfrak M$ at $\vev\Sigma$.
The Cartan-Weyl basis has a nice property,\footnote{
Since the algebra is invariant under a similarity transformation, one can also see it in the original basis
\eqref{eq:Sigma basis},
$[  H_i, E_\alpha] = \vec\alpha_i  E_\alpha$.
}
\begin{equation}
\label{eq:Sigma basistilde}
    [ \tilde H_i,\tilde E_\alpha] = \vec\alpha_i \tilde E_\alpha \, , 
\end{equation}
for $\alpha = \pm t, \pm u, \pm v$ and $i=0,1,2$.
Here, $\vec \alpha_i$ is the root of the ladder operator $\tilde E_\alpha$,
whose components are listed in the following table:
\begin{equation}
    \begin{array}{c|ccc}
        \alpha & \vec\alpha_0 & \vec\alpha_1 & \vec\alpha_2\\
        \hline
        {\pm t} & 0 & \pm \sqrt2 & 0\\
        {\pm u} & 0 & \mp \frac1{\sqrt2} & \pm\sqrt{\frac32}\\
        {\pm v} & 0 & \pm \frac1{\sqrt2} & \pm\sqrt{\frac32}
    \end{array} 
\end{equation}
Eq.~\eqref{eq:Sigma basistilde} implies
\begin{equation}
    D_{\alpha} \Sigma =  \left(\vec \alpha_i \vev\Sigma^i_{\rm diag}\right) \tilde E_\alpha \, .\label{eq:Lie derivative}
\end{equation}
These Lie derivatives can be regarded as the tangent vectors of $\mathfrak M$ at $\vev\Sigma$.

Since we have found a coordinate system for $\mathfrak M$ around $\vev\Sigma$ and
identified the tangent space at $\vev\Sigma$ given by its Lie derivatives,
let us now consider stabilizing the moduli space $\mathfrak M$.
To stabilize all degrees of freedom in $\mathfrak M$ around the vacuum,
the superpotential of Eq.~\eqref{Weffmoduli} must satisfy the following condition at $\langle \Sigma \rangle$:
\begin{equation}
    \dfrac{\partial^2 W_{\rm eff}}{\partial \theta^\alpha \partial \theta^\beta} > 0  \, .  \label{eq:F term stability}
\end{equation}
Here, $>0$ represents positive-definiteness, and $\alpha, \beta$ run over $\pm t, \pm u, \pm v$
which correspond to the bases of the tangent space of $\mathfrak M$ at $\vev\Sigma$.
In \SEC{sec:vacuum}, we have shown that $W_{\rm eff} = W_0 + \delta W + {\rm const.}$
whose $W_0$ part is immune to similarity transformations, and hence only the contribution from $\delta W$ needs to be considered,
\begin{equation}
    \dfrac{\partial^2 \delta W}{\partial \theta^\alpha \partial \theta^\beta} > 0  \, .
\end{equation}
Since $\delta W$ in Eq.~\eqref{eq:delta W} reveals a symmetry, $\lambda_{ik}^{~jl} = \lambda_{ki}^{~lj}$, 
there are $n_{\Sigma}^2(n_{\Sigma}^2+1)/2$ independent tensor elements in $\lambda_{ik}^{~jl}$.
Under the basis $\{\tilde \sigma_a \}$, the coupling $\lambda$ and superpotential $\delta W$ can be decomposed as
\begin{eqnarray}
     \lambda &=&  \lambda^{ab} \tilde\sigma_a \otimes \tilde\sigma_b \, ,\quad
    \lambda^{ab} = \lambda^{ba} \, ,\\[1ex]
    \delta W &=&  \frac{\Lambda^2}{\mu} \lambda^{a'b'}g_{aa'}g_{bb'} \left( \Sigma^a - \langle\Sigma\rangle^a\right)\left( \Sigma^b - \langle\Sigma\rangle^b\right)  , \nonumber\\ \label{eq:delta W eff expansion}
\end{eqnarray}
with $a,a',b,b'$ running from $0$ to $n_\Sigma^2-1=8$.
Then we can calculate the second-order derivatives of $\delta W$ according to Eq.~\eqref{eq:SL transformation} and Eq.~\eqref{eq:Lie derivative}.
For $\alpha,\beta =\pm t, \pm u, \pm v$, we have
\begin{equation}
    \dfrac{\partial^2 \delta W}{\partial \theta^\alpha \partial \theta^\beta} 
    = -\dfrac{2\Lambda^2}{\mu} \lambda^{\alpha' \beta'} g_{\alpha \alpha'}g_{\beta' \beta} (\vec \alpha_i \vev\Sigma^i_{\rm diag}) (\vec \beta_j \vev\Sigma^j_{\rm diag}) \, ,
    \label{Wcondition}
\end{equation}
where $\vec \alpha_i$ and $\vec \beta_j$ denote the roots corresponding to $E_\alpha$ and $E_\beta$, respectively,
and
$\vev \Sigma^i_{\rm diag}$ are the components of $\vev\Sigma_{\rm diag}$ expanded in the Cartan subalgebra basis $H_{0,1,2}$.
To make the matrix \eqref{Wcondition} positive-definite,
as an example, one can set
\begin{eqnarray}
    \vev\Sigma = \left(
    \begin{array}{ccc}
        \Sigma_1 & \frac{i}2 \Sigma_2 \cos\phi & 0\\
        -\frac{i}2 \Sigma_2 \cos\phi & \Sigma_1 + \Sigma_2 \sin\phi & 0\\
        0 & 0 & \Sigma_3
    \end{array}
    \right),\label{eq:VEV example}
\end{eqnarray}
with $\Sigma_1,\Sigma_2,\Sigma_3$ and $\phi$ real constants,
and take the non-vanishing components of $\lambda$ as
\beq
&&\lambda^{00} = \lambda_1 \, , \quad \lambda^{11} = - \lambda_2 \, , \quad \lambda^{22} = \lambda_3 \, ,\nonumber\\[1.5ex]
&&\lambda^{tt} = - \lambda_2\left(1-\frac{2}{\sin\phi-1}\right) , \nonumber \\
&&\lambda^{-t,-t} = \lambda_2\left(1-\frac{2}{\sin\phi+1}\right) ,\nonumber\\[1ex]
&&\lambda^{uu} = - \lambda_4  \left(1-\sin\phi\right)\Sigma_2/\Lambda \, , \\[3ex]
&&\lambda^{-u,-u} = -\lambda_5 (1+\sin\phi) \Sigma_2/\Lambda \, , \nonumber\\[3ex]
&&\lambda^{vv} = -\lambda_4  \left(1+\sin\phi\right)\Sigma_2/\Lambda \, , \nonumber\\[3ex]
&&\lambda^{-v,-v} = -\lambda_5  (1-\sin\phi)\Sigma_2/\Lambda \, ,\nonumber
\eeq
where $\lambda_1,\cdots , \lambda_5$ are real constants satisfying $\lambda_2$, $\lambda_4 \Sigma_2$, $\lambda_5 \Sigma_2>0$,
and we choose the Cartan-Weyl basis, $\{\tilde\sigma_a\}=\{U^{-1}_0 H_{0,1,2} U_0, U^{-1}_0 E_{\pm t, \pm u, \pm v} U_0\}$ with
a similarity transformation that diagonalizes $\vev\Sigma$,
\begin{equation}
    U_0 = \left(
    \begin{array}{ccc}
        i(\tan\phi - \sec\phi) & 1 & 0\\
        i(\tan\phi + \sec\phi) & 1 & 0\\
        0 & 0 & 1
    \end{array}
    \right) .
\end{equation}
With this choice of the parameters, one can check that the matrix \eqref{Wcondition} is positive-definite.
In fact, according to Eq.~\eqref{eq:diag Sigma}, 
the current $\vev\Sigma$ leads to
\beq
\vev\Sigma^i_{\rm diag} &=& \langle \vev\Sigma, U^{-1}_0 H_i U_0 \rangle\nonumber\\
&=& \frac{1}{\sqrt6}\left( 
\begin{array}{c}
    \sqrt2 \left(2\Sigma_1 +  \Sigma_3 + \Sigma_2\sin\phi \right) \\
    -\sqrt3 \Sigma_2\\
     2\Sigma_1 - 2\Sigma_3 + \Sigma_2 \sin\phi
\end{array}
 \right),~~~~~~~
\eeq
where $i$ runs over $0,1,2$.
The $\dfrac{\partial^2 \delta W}{\partial \theta^\alpha \partial \theta^\beta}$ is diagonal and its non-vanishing components 
are given by
\beq
&&\dfrac{\partial^2 \delta W}{\partial \theta^t \partial \theta^t} = \dfrac{2\Lambda^2}{\mu} \lambda_2 \left(\frac{2}{s_\phi+1}-1\right) \left(\Sigma_2\right)^2,\nonumber\\[1.5ex]
&&\dfrac{\partial^2 \delta W}{\partial \theta^{-t} \partial \theta^{-t}} = \dfrac{2\Lambda^2}{\mu} \lambda_2 \left(1+ \frac{2}{1- s_\phi}\right) \left(\Sigma_2\right)^2,\nonumber\\[1.5ex]
&&\dfrac{\partial^2 \delta W}{\partial \theta^u \partial \theta^u} = \dfrac{2\Lambda}{\mu} \lambda_5 \left(1+s_\phi\right) \Sigma_2 \left(\Sigma_1 - \Sigma_3 +  \frac{1+s_\phi}{2}\Sigma_2 \right)^2, \nonumber\\[1.5ex]
&&\dfrac{\partial^2 \delta W}{\partial \theta^{-u} \partial \theta^{-u}} = \dfrac{2\Lambda}{\mu} \lambda_4 \left(1-s_\phi\right) \Sigma_2 \left(\Sigma_1 - \Sigma_3 +  \frac{1+s_\phi}{2}\Sigma_2 \right)^2,\nonumber\\[1.5ex]
&&\dfrac{\partial^2 \delta W}{\partial \theta^v \partial \theta^v} = \dfrac{2\Lambda}{\mu} \lambda_5 \left(1-s_\phi\right) \Sigma_2 \left(\Sigma_1 - \Sigma_3 -  \frac{1-s_\phi}{2}\Sigma_2 \right)^2,\nonumber\\[1.5ex]
&&\dfrac{\partial^2 \delta W}{\partial \theta^{-v} \partial \theta^{-v}} = \dfrac{2\Lambda}{\mu} \lambda_4 \left(1+s_\phi\right) \Sigma_2 \left(\Sigma_1 - \Sigma_3 -  \frac{1-s_\phi}{2}\Sigma_2 \right)^2, \nonumber \\
{}
\eeq
where $s_\phi$ is a shorthand for $\sin\phi$.
The above expressions show positive-definiteness of the second-order derivatives of the superpotential.
Therefore, the moduli space $\mathfrak M$ is stabilized at $\vev\Sigma$.

Note that $\det\vev\Sigma\in \mathbb{R}$ and $\det\vev\Sigma\ne 0$
when $\Sigma_3\ne 0$ and $\Sigma_1$ equals neither $\Sigma_2(\pm 1 \pm \sin\phi)/2$ nor $\Sigma_2(\pm 1 \mp \sin\phi)/2$.
Since $\im \vev\Sigma^\dagger \vev\Sigma = [\im \vev\Sigma, \re \vev\Sigma]\ne 0$,
the CKM phase is correctly generated.
The coupling $\lambda_{ik}^{~jl}$ is real so that the $\lambda \Sigma \Sigma$ term in the superpotential is CP-conserving.
This can be seen by going to the basis $\{\sigma_a\}=\{H, E\}$.
Since $\{H, E\}$ is a real basis, if we can confirm that $\lambda^{ab}$ is real in this basis,
then $\lambda_{ik}^{~jl}$ is real.
In the basis $\{\sigma_a\}=\{H, E\}$, the non-vanishing elements of $\lambda^{ab}$ are given by
\beq
&&\lambda^{00} = \lambda_1 \, , \quad \lambda^{11} = -\lambda_2 \, ,
\quad \lambda^{22} = \lambda_3 \, , \nonumber\\[1.5ex]
&&\lambda^{\pm t \pm t} = \lambda_2 \, , \nonumber\\[1.5ex]
&&\lambda^{uu} = -\lambda^{vv} = -\lambda_4  \Sigma_2/(2\Lambda)\cos^2\phi \, , \nonumber\\[1.5ex]
&&\lambda^{-v-v} = -\lambda^{-u-u} = 2\lambda_5\Sigma_2/\Lambda \, ,
\eeq
which are all real.
The mass parameter $M$ in Eq.~\eqref{eq:Mass} can be also real.
Note that the imaginary part of the adjugate matrix $\tilde{\vev\Sigma}$ only comes from its off-diagonal components,
\begin{widetext}
\beq
\tilde{\vev\Sigma} &=& \left(
\begin{array}{ccc}
\Sigma_1\Sigma_3 + \Sigma_2\Sigma_3 s_\phi & -\frac{i}2 \Sigma_2 \Sigma_3 c_\phi & 0\\
\frac{i}2 \Sigma_2 \Sigma_3 c_\phi & \Sigma_1 \Sigma_3 & 0\\
0 & 0 & \Sigma_1^2 - \Sigma_2^2 c_\phi^2 /4 + \Sigma_1\Sigma_2 s_\phi
\end{array}\right) .
\eeq
\end{widetext}
Then, the imaginary part can be canceled by the second term of Eq.~\eqref{eq:Mass} 
as long as $\Sigma_3 = \lambda_2 \frac{\Lambda^5}{(4\pi)^4\mu c \Phi^3}$ is satisfied.
Therefore, the parameter $M$ preserves CP symmetry.

\section{UV completion
\label{app:UV}}

In our model, we consider some higher-dimensional operators in \EQ{UV} and \EQ{Yukawa} for inducing spontaneous CP violation and reproducing the Yukawa interactions of the down-type quarks.
Here, we discuss a UV completion of those higher-dimensional operators.
The superpotential can be given by
\beq
W_{\rm UV} &=& -m^{(X)} X_p^{~q} X_q^{~p} +\lambda^{(X)} X_q^{~p} \bar{Q}^{qn}Q_{pn}\nonumber\\
&-& \sum_{ijkl} m_{ijkl}^{(Y)}Y_{ij}Y_{kl} +\sum_{ij}\lambda_{ij}^{(Y)}Y_{ij}\bar{Q}^{in}Q_{jn}\nonumber\\
&-&\sum_{ij}m_{ij}^{(Z)}Z_{in} \bar{Z}^{jn}\nonumber\\
&+& \sum_i \lmk \lambda_i^{(Z)} q_p^i Z_{in} \bar{Q}^{pn} + \bar{\lambda}_i^{(Z)} \bar{Z}^{in} Q_{in} H_d\rmk,~~~
\label{uv}
\eeq
where we have introduced new $SU(5)$ gauge singlet fields $X_p^{~q},Y_{ij}$ and a vector-like pair $Z_i,\bar{Z}^i$
transforming as the (anti-)fundamental representation.
Table \ref{tab:UV} summarizes the charge assignment.
Their mass parameters are denoted by $m^{(X)},m^{(Y)},m^{(Z)}$, and the coupling coefficients are $\lambda^{(X)}, \lambda^{(Y)}, \lambda^{(Z)}$.
They are all real due to CP symmetry.

In this setup, the number of vector-like flavors for the $SU(5)$ gauge theory is $6+6=12$,
and the theory is in conformal window \cite{Intriligator:2007cp}.
At the scale of $m^{(Z)}$, $Z,\bar{Z}$ are integrated out,
and the effective theory shows s-confinement as we have seen in the main text.
This scale corresponds to $\mu'$ in \EQ{UVyukawa}, and the dynamical scale $\Lambda$ is naturally set to be right below
the scale.
The masses $m^{(X)},m^{(Y)}$ can be arbitrary scales higher than $\Lambda$.
Integrating out the heavy fields, we obtain the higher-dimensional operators in \EQ{UV} and \EQ{Yukawa}.

\begin{table}[!h]
\vspace{0mm}
\centering
\begin{tabular}{c|c|c|c|c}
 & $X_p^{~q}$ & $Y_{ij}$ & $Z_i$ & $\bar{Z}^i$  \\
\hline
$SU(5)$ & ${\bf 1}$ & ${\bf 1}$ & $\square$ & $\overline{\square}$  \\
$SU(3)_c$ & ${\bf 1}+${\bf Ad} & ${\bf 1}$ & $\bf{1}$ & $\bf{1}$ \\
$SU(2)_L$ & ${\bf 1}$ & ${\bf 1}$ & $\bf 2$ & $\bf 2$ \\
$U(1)_Y$ & 0 & 0 & $-3/10$ & $3/10$  \\
\end{tabular}
\vspace{1mm}
\caption{The charge assignment of a UV completed model.}
\label{tab:UV}
\end{table}

\bibliography{reference}

\end{document}